\documentclass[sn-standardnature,iicol]{sn-jnl}

\usepackage{siunitx}
\usepackage{amssymb}
\usepackage{times}
\usepackage{booktabs}
\usepackage{amsmath, array, multirow, rotating, makecell}

\usepackage{xcolor}%

\newcommand\aj{{Astron.~J.}}
\newcommand\apj{{Astrophys.~J.}}
\newcommand\apjl{{Astrophys.~J.~Lett.}}

\newcommand\aap{{Astron.~Astrophys.}}
\newcommand\mnras{{Mon.~Not.~R.~Astron.~Soc.}}
\newcommand\nat{{Nature}}

\newcommand\pasp{{Publ. Astron. Soc. Pac.}}

\jyear{2023}%

\begin{document}

\title{\centering {Inhomogeneous terminators on the exoplanet WASP-39~b}}

%%=============================================================%%
%% Prefix	-> \pfx{Dr}
%% GivenName	-> \fnm{Joergen W.}
%% Particle	-> \spfx{van der} -> surname prefix
%% FamilyName	-> \sur{Ploeg}
%% Suffix	-> \sfx{IV}
%% NatureName	-> \tanm{Poet Laureate} -> Title after name
%% Degrees	-> \dgr{MSc, PhD}
%% \author*[1,2]{\pfx{Dr} \fnm{Joergen W.} \spfx{van der} \sur{Ploeg} \sfx{IV} \tanm{Poet Laureate} 
%%                 \dgr{MSc, PhD}}\email{iauthor@gmail.com}
%%=============================================================%%

\author[1,2]{\fnm{N\'estor} \sur{Espinoza}} %\email{nespinoza@stsci.edu}
\author[3,4]{\fnm{Maria} E. \sur{Steinrueck}}
\author[5]{\fnm{James} \sur{Kirk}}
\author[6]{\fnm{Ryan} J. \sur{MacDonald}}
\author[7]{\fnm{Arjun} B. \sur{Savel}}
\author[7]{\fnm{Kenneth} \sur{Arnold}}
\author[7]{\fnm{Eliza}  M.-R.  \sur{Kempton}}
\author[8]{\fnm{Matthew} M. \sur{Murphy}}
\author[9]{\fnm{Ludmila} \sur{Carone}}
\author[10]{\fnm{Maria} \sur{Zamyatina}}
\author[9,11]{\fnm{David} A. \sur{Lewis}}
\author[9]{\fnm{Dominic} \sur{Samra}}
\author[12,13,9,14]{\fnm{Sven} \sur{Kiefer}}
\author[6]{\fnm{Emily} \sur{Rauscher}}
\author[4]{\fnm{Duncan} \sur{Christie}}
\author[10]{\fnm{Nathan} \sur{Mayne}}
\author[9,11]{\fnm{Christiane} \sur{Helling}}
\author[2]{\fnm{Zafar} \sur{Rustamkulov}}
\author[15]{\fnm{Vivien} \sur{Parmentier}}
\author[16]{\fnm{Erin} M. \sur{May}}
\author[17]{\fnm{Aarynn} L. \sur{Carter}}
\author[18]{\fnm{Xi} \sur{Zhang}}
\author[19]{\fnm{Mercedes} \sur{L\'opez-Morales}}
\author[2]{\fnm{Natalie} \sur{Allen}}
\author[20,21]{\fnm{Jasmina} \sur{Blecic}}
\author[12]{\fnm{Leen} \sur{Decin}}
% something weird happens with Mancini's when ingesting
% [22,23,3], so put the 22 by hand on the name:
\author[23,3]{\fnm{Luigi} \sur{Mancini}$^{\textrm{22,}}$}
\author[24,25]{\fnm{Karan} \sur{Molaverdikhani}}
\author[26,27]{\fnm{Benjamin} V. \sur{Rackham}}
\author[28]{\fnm{Enric} \sur{Palle}}
\author[29]{\fnm{Shang-Min} \sur{Tsai}}
\author[30]{\fnm{Eva-Maria} \sur{Ahrer}}
\author[4]{\fnm{Jacob} L. \sur{Bean}}
\author[31]{\fnm{Ian} J. M. \sur{Crossfield}}
\author[3]{\fnm{David} \sur{Haegele}}
\author[10]{\fnm{Eric} \sur{H\'ebrard}}
\author[3]{\fnm{Laura} \sur{Kreidberg}}
\author[4]{\fnm{Diana} \sur{Powell}}
\author[12,32]{\fnm{Aaron} D. \sur{Schneider}}
% Same with Welbanks:
\author[33]{\fnm{Luis} \sur{Welbanks}$^{\textrm{33}}$}
\author[30]{\fnm{Peter} \sur{Wheatley}}
\author[34,35,36]{\fnm{Rafael} \sur{Brahm}}
\author[37]{\fnm{Nicolas} \sur{Crouzet}}

\affil[1]{\orgname{Space Telescope Science Institute}, \orgaddress{\street{3700 San Martin Drive}, \city{Baltimore}, \postcode{21218}, \state{MD}, \country{USA}}}

\affil[2]{\orgdiv{Department of Physics \& Astronomy}, \orgname{Johns Hopkins University}\orgaddress{, \city{Baltimore}, \postcode{21218}, \state{MD}, \country{USA}}}

\affil[3]{\orgname{Max Planck Institute for Astronomy (MPIA)}, \orgaddress{\street{K\"onigstuhl 17}, \city{Heidelberg}, \postcode{D-69117}, \country{Germany}}}

\affil[4]{\orgdiv{Department of Astronomy \& Astrophysics}, \orgname{University of Chicago}, \orgaddress{\city{Chicago}, \state{IL} \country{USA}}}

\affil[5]{\orgdiv{Department of Physics}, \orgname{Imperial College London}, \orgaddress{\street{Prince Consort Road}, \city{London}, \postcode{SW7 2AZ}, \country{UK}}}

\affil[6]{\orgdiv{Department of Astronomy}, \orgname{University of Michigan}, \orgaddress{\street{1085 S. University Ave.}, \city{Ann Arbor}, \postcode{48109}, \state{MI}, \country{USA}}}

\affil[7]{\orgdiv{Department of Astronomy}, \orgname{University of Maryland,}, \orgaddress{\city{College Park}, \postcode{20742}, \state{MD}, \country{USA}}}

\affil[8]{\orgdiv{Department of Astronomy and Steward Observatory}, \orgname{University of Arizona}, \orgaddress{\street{933 North Cherry Avenue}, \city{Tucson}, \postcode{85721}, \state{AZ}, \country{USA}}}

\affil[9]{\orgdiv{Space Research Institute}, \orgname{Austrian Academy of Sciences}, \orgaddress{\street{Schmiedlstrasse 6}, \city{Graz}, \postcode{A-8042}, \country{Austria}}}

\affil[10]{\orgdiv{Department of Physics and Astronomy, Faculty of Environment, Science and Economy}, \orgname{University of Exeter}, \orgaddress{\city{Exeter}, \postcode{EX4 4QL}, \country{UK}}}

\affil[11]{\orgdiv{Institute for Theoretical Physics and Computational Physics}, \orgname{Graz University of Technology}, \orgaddress{\street{Petersgasse 16}, \city{Graz}, \country{Austria}}}

\affil[12]{\orgdiv{Institute of Astronomy}, \orgname{KU Leuven}, \orgaddress{\street{Celestijnenlaan 200D}, \city{Leuven}, \postcode{3001}, \country{Belgium}}}

\affil[13]{\orgdiv{Centre for Exoplanet Science}, \orgname{University of St Andrews}, \orgaddress{\street{North Haugh}, \city{St Andrews}, \postcode{KY169SS}, \country{UK}}}

\affil[14]{\orgdiv{Fakult\"at f\"ur Mathematik, Physik und Geod\"asie}, \orgname{TU Graz}, \orgaddress{\street{Petersgasse 16}, \city{Graz}, \postcode{A-8010}, \country{Austria}}}

\affil[15]{\orgdiv{Observatoire de la C\^ote d'Azur}, \orgname{Universit\'e C\^ote d'Azur, CNRS}, \orgaddress{\street{Laboratoire Lagrange}, \city{Nice}, \country{France}}}

\affil[16]{\orgname{Johns Hopkins APL,}, \orgaddress{\street{11100 Johns Hopkins Rd}, \city{Laurel}, \postcode{20723}, \state{MD}, \country{USA}}}

\affil[17]{\orgdiv{Department of Astronomy \& Astrophysics}, \orgname{University of California, Santa Cruz}, \orgaddress{\street{1156 High St}, \city{Santa Cruz}, \postcode{95064}, \state{CA}, \country{USA}}}

\affil[18]{\orgdiv{Department of Earth and Planetary Sciences}, \orgname{University of California, Santa Cruz}, \orgaddress{\street{1156 High St}, \city{Santa Cruz}, \postcode{95064}, \state{CA}, \country{USA}}}

\affil[19]{\orgdiv{Center for Astrophysics}, \orgname{Harvard \& Smithsonian}, \orgaddress{\street{60 Garden Street}, \city{Cambridge}, \postcode{02138}, \state{MA}, \country{USA}}}

\affil[20]{\orgdiv{Department of Physics}, \orgname{New York University Abu Dhabi}, \orgaddress{\street{PO Box 129188 Abu Dhabi}, \country{UAE}}}

\affil[21]{\orgdiv{Center for Astro, Particle, and Planetary Physics (CAP3)}, \orgname{New York University Abu Dhabi}, \orgaddress{\street{PO Box 129188 Abu Dhabi}, \country{UAE}}}

\affil[22]{\orgdiv{Department of Physics}, \orgname{University of Rome ``Tor Vergata"}, \orgaddress{\city{Rome}, \country{Italy}}}

\affil[23]{\orgname{INAF - Turin Astrophysical Observatory}, \orgaddress{\city{Pino Torinese}, \country{Italy}}}

\affil[24]{\orgdiv{University Observatory Munich}, \orgname{Ludwig Maximilian University}, \orgaddress{\city{Munich}, \country{Germany}}}

\affil[25]{\orgname{Exzellenzcluster Origins}, \orgaddress{\city{Garching}, \country{Germany}}}

\affil[26]{\orgdiv{Department of Earth, Atmospheric and Planetary Sciences}, \orgname{Massachusetts Institute of Technology}, \orgaddress{\city{Cambridge}, \state{MA}, \country{USA}}}

\affil[27]{\orgdiv{Kavli Institute for Astrophysics and Space Research}, \orgname{Massachusetts Institute of Technology}, \orgaddress{\city{Cambridge}, \state{MA}, \country{USA}}}

\affil[28]{\orgname{Instituto de Astrofísica de Canarias (IAC)}, \orgaddress{\city{Tenerife}, \country{Spain}}}

\affil[29]{\orgdiv{Department of Earth Sciences}, \orgname{University of California, Riverside}, \orgaddress{\city{Riverside}, \state{CA}, \country{USA}}}

\affil[30]{\orgdiv{Centre for Exoplanets and Habitability}, \orgname{University of Warwick}, \orgaddress{\city{Coventry}, \country{UK}}}

\affil[31]{\orgdiv{Department of Physics \& Astronomy}, \orgname{University of Kansas}, \orgaddress{\city{Lawrence}, \state{KS} \country{USA}}}

\affil[32]{\orgdiv{Centre for ExoLife Sciences}, \orgname{Niels Bohr Institute}, \orgaddress{\city{Copenhagen}, \country{Denmark}}}

\affil[33]{\orgdiv{School of Earth and Space Exploration}, \orgname{Arizona State University}, \orgaddress{\city{Tempe}, \state{AZ} \country{USA}}}

\affil[34]{\orgdiv{Facultad de Ingenier\'ia y Ciencias}, \orgname{Universidad Adolfo Ib\'a\~nez}, \orgaddress{\street{Av. Diagonal las Torres 2640}, \city{Santiago}, \country{Chile}}}

\affil[35]{\orgname{Millennium Institute for Astrophysics}, \orgaddress{\street{Av. Vicu\~na Mackenna 4860}, \city{Santiago}, \country{Chile}}}

\affil[36]{\orgname{Data Observatory Foundation}, \orgaddress{\street{Eliodoro Y\'a\~nez 2990}, \city{Santiago}, \country{Chile}}}

\affil[37]{\orgdiv{Leiden Observatory}, \orgname{Leiden University}, \orgaddress{\street{P.O. Box 9513, 2300 RA}, \city{Leiden}, \country{The Netherlands}}}

\maketitle

\textbf{{Transmission spectroscopy has been a workhorse technique over the past two decades to constrain the physical and 
chemical properties of exoplanet atmospheres \cite{seager2000, hubbard2001, burrows2003, 
fortney2005, kreidberg:2018}. One of its classical key assumptions is that the portion of the 
atmosphere it probes --- the terminator region --- is homogeneous. Several works in the past decade, however, have put this into question for highly irradiated, hot ($T_{eq}\gtrsim 1000$ K) gas 
giant exoplanets both empirically \cite{LW:2015, ehrenreich:2020, prinoth:2022, Kesseli:2022, prism:2023} and via 3-dimensional modelling \cite{fortney:2010, dd:2012, lp:2016, kempton:2017, powell:2019, helling:2020, macdonald:2020}. While models predict clear differences between the evening (day-to-night) and morning (night-to-day) terminators, \textit{direct} morning/evening transmission spectra in a wide wavelength range has not been reported for an exoplanet to date. {Under the assumption of precise and accurate orbital parameters on WASP-39~b,} here we report the detection of inhomogeneous terminators on the exoplanet WASP-39~b, which allows us to retrieve its morning and evening transmission spectra in the near-infrared ($2-5\ \mu$m) using \textit{JWST}. We observe larger transit depths in the evening which are, on average, $405 \pm 88$ ppm larger than the morning ones}, also having qualitatively larger features than the morning spectrum. The spectra are best explained by models in which the evening terminator is hotter than the morning terminator by $177^{+65}_{-57}$ K with both terminators having C/O ratios consistent with solar. General circulation models (GCMs) predict temperature differences broadly consistent with the above value and point towards a cloudy morning terminator and a clearer evening terminator.}


%%\pacs[JEL Classification]{D8, H51}

%%\pacs[MSC Classification]{35A01, 65L10, 65L12, 65L20, 65L70}

\maketitle

Our study was performed using observations of WASP-39~b from the \textit{JWST} Transiting 
Exoplanet Community Director’s Discretionary Early Release Science (ERS) program (ERS-1366; PIs: N. M. Batalha, J. L. Bean, K. B. Stevenson) \cite{stevensonERS, beanERS}. %, which targeted observations of the exoplanet WASP-39~b. 
This highly irradiated gas giant 
exoplanet has a mass of 0.28$M_{Jup}$, a radius of $1.27R_{Jup}$, and an equilibrium temperature of 1100 K. The ERS observations consisted of four transit events observed 
with four different \textit{JWST} instruments/modes, which combined reveal prominent atomic and molecular absorption features in the terminator region, 
including K, H$_2$O, CO$_2$, CO and even SO$_2$, which was identified as a 
photochemical product \cite{aldersonERS, feinsteinERS, ahrer:2023, prism:2023, tsai:2023}. Our analysis is performed in particular on the NIRSpec/PRISM observations of WASP-39~b \cite{prism:2023}, as this dataset has the widest wavelength coverage while simultaneously presenting minimal instrumental systematics (see Carter \& May et al., in review). This consists of an 8.23-hour observation centered around the 10 July 2022 transit of WASP-39~b.

The \textit{JWST} data were reduced using the \texttt{FIREFLy} pipeline \cite{prism:2023} as described in the work of Carter \& May et al. (in review). {Carter \& May et al. (in review) also demonstrate that the majority of wavelengths less than 2 $\mu$m suffer from detector saturation, and are not in agreement with measurements performed using the NIRISS SOSS instrument across a similar wavelength range \cite{feinsteinERS}. As a reliable determination of the transit depth $<$2~$\mu$m cannot be obtained using the NIRSpec PRISM observations, we opt to use only the 2-5 $\mu$m data in our present analysis.} We fit each individual wavelength-dependent light curve at the pixel-level resolution element of 
the instrument, using a simple linear term in time 
as our systematics model, following the work of \cite{prism:2023}. 

\begin{figure}
    \centering
    \includegraphics[width=\linewidth]{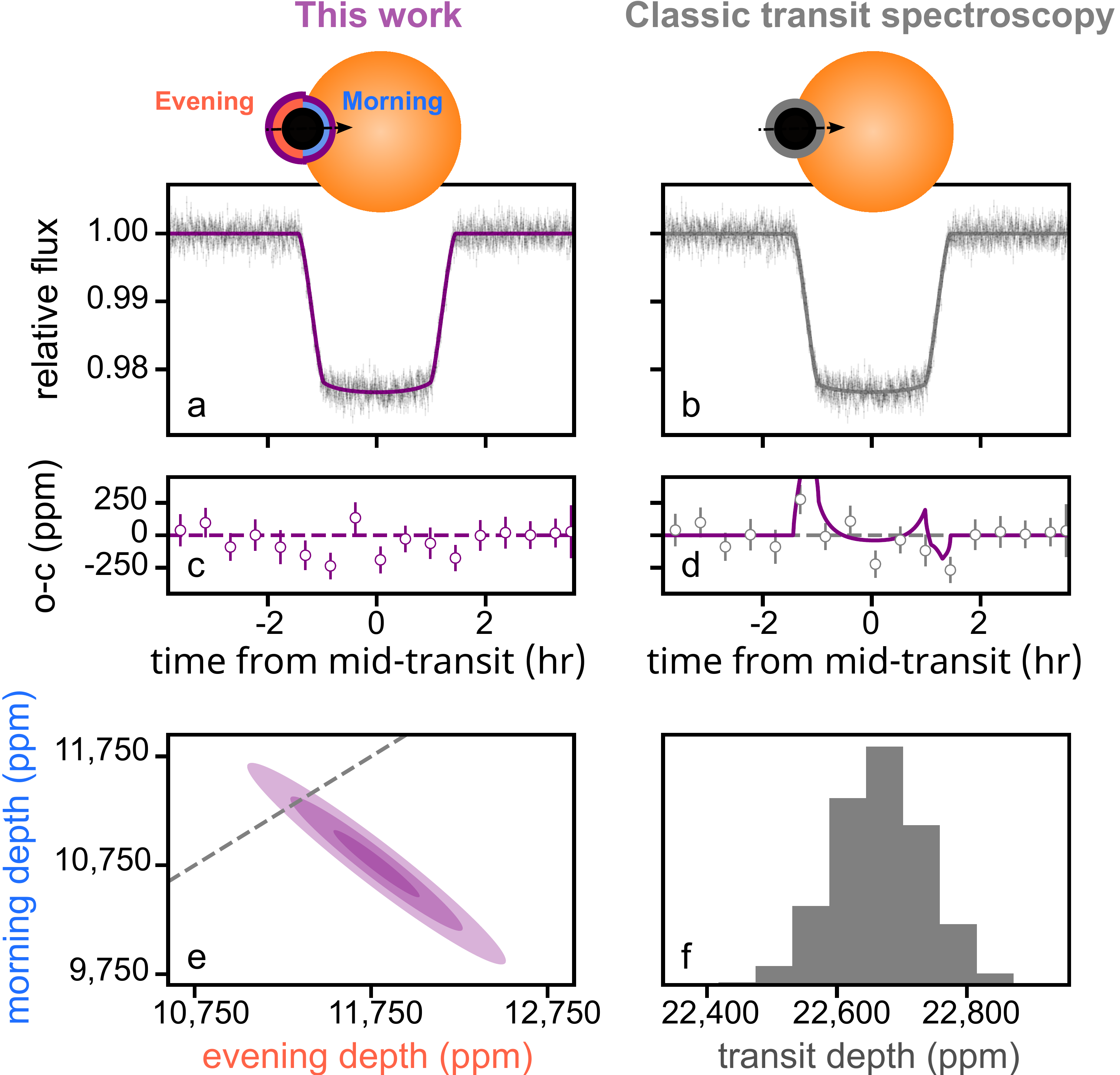}
    \caption{\textbf{Light curve modeling and extraction of morning and evening depths.} \textbf{a-b.} Transit lightcurve modelling of the July 10th, 2022 transit at 4.38 microns {(at the middle of the CO$_2$ spectral feature in the transmission spectrum) }of 
    WASP-39~b (grey datapoints) using both the \texttt{catwoman} framework (\textbf{a}, purple 
    model) and the classic circular occulter model via \texttt{batman} (\textbf{b}, grey model). \textbf{c-d.} Residuals of the best-fit model for each methodology. The \texttt{batman} residuals (panel \textbf{d}) show, in 
    turn, the difference between the \texttt{catwoman} and \texttt{batman} light curve models (purple line), showcasing how the \texttt{catwoman} models small ($\sim$ few 100s of ppm) light curve asymmetries. Note the 
    residuals are of comparable magnitude; while it is difficult to conclude the effect is 
    present in each individual lightcurve, the effect is detectable once all the wavelength-dependent lightcurves have been analyzed (see Figure \ref{fig:morning-evening-spectra} 
    and text for details). \textbf{e-f.} Inferences enabled from each methodology. The 
    \texttt{catwoman} methodology (\textbf{e}) allows to extract morning/evening 
    transit depths (purple ellipses representing the 1, 2 and 3-sigma posterior contours; 
    dashed grey line indicates equal morning and evening depths); the circular occulter methodology (\textbf{f}) only allows to extract a single total 
    transit depth from the lightcurve. {All errorbars represent 1-standard deviation.}}
    \label{fig:morning-evening-lc} 
\end{figure}

The transit light curve analysis was performed via three different methodologies, all of which entail different assumptions yet provide consistent results (see Methods for details). We report on the results obtained using the \texttt{catwoman} framework \cite{cw:2020, EJ:2021}, which 
is arguably the most conservative of the approaches{ as it allows to fit for morning and evening terminators simultaneously in a single fit}. The framework models the terminator as two semi-circles of independent radii, thus allowing us to separately retrieve the sizes of both morning and evening limbs as a function of wavelength from the transit light curves themselves. An example lightcurve fit at $4.38 \mu$m using this methodology, as well as the classic circular occulter method --- here modeled using the \texttt{batman} library 
\cite{batman} --- is presented in Figure \ref{fig:morning-evening-lc}. More details on the data analysis are given in the Methods section.

\begin{figure}
    \centering
    \includegraphics[width=\linewidth]{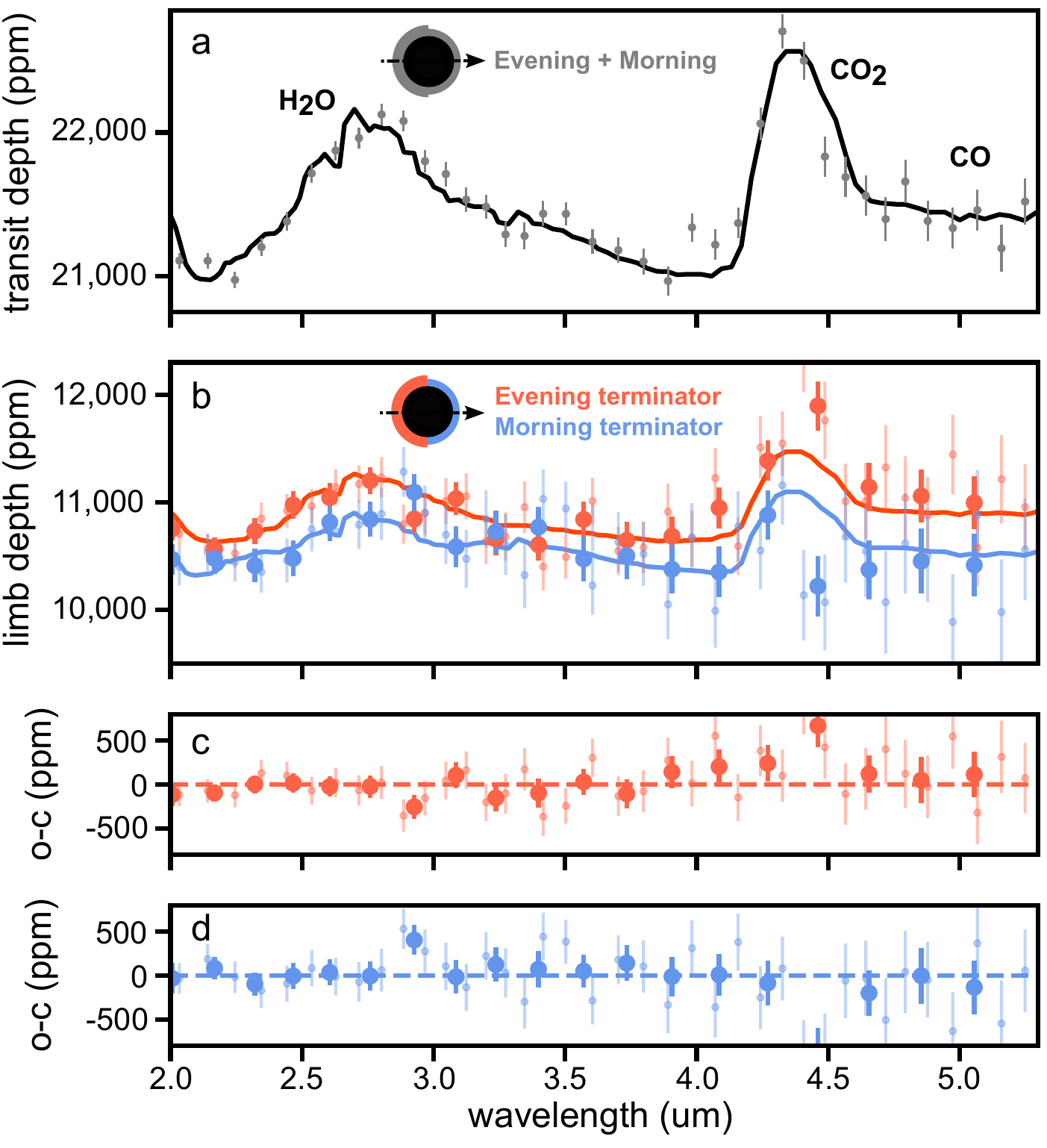}
    \caption{\textbf{The morning and evening spectrum of WASP-39~b from JWST NIRSpec/PRISM observations.} \textbf{a-b.} The total transit depth by adding the morning and evening spectra (\textbf{a}, grey points) along with the individual morning (blue points) and evening (red points) spectrum of WASP-39~b as derived from our lightcurve modeling (\textbf{b}). Big points are datapoints at $R=30$ shown for illustration; smaller points are at $R=100$. The best-fit, chemically-consistent models (black solid line in \textbf{a}, red and blue solid lines in \textbf{b}; fitted to the $R=100$ spectra) are consistent with a hotter evening terminator (see text for details). \textbf{c-d.} Residuals from the best-fit model (red points for the evening, blue points for the morning); dashed line marks 0. {All errorbars represent 1-standard deviation.}}
    \label{fig:morning-evening-spectra} 
\end{figure}

The resulting morning and evening terminator spectra inferred from our \texttt{catwoman} 
lightcurve fits are presented in Figure \ref{fig:morning-evening-spectra}b. We compute 
an average difference between the evening and morning spectra of $405 \pm 88$ ppm, 
finding this difference to be inconsistent with 0 at more than $4.6\sigma$ --- i.e., the spectra show 
statistically distinct features in the morning and evening terminators. Qualitatively, both the morning and evening spectra showcase transit depth increases at around the H$_2$O ($2.84\ \mu$m) and CO$_2$ ($4.38\ \mu$m) features, with the morning spectra being consistent with somewhat flatter spectra --- a behavior we observe with all our light curve fitting methodologies. {This detection of evening-to-morning spectral differences is, in turn, robust at the $3\sigma-$level even when accounting for the current best uncertainties on the orbital parameters of WASP-39~b. The result does, 
however, depend on the accuracy and precision of those parameters up to a factor of a few of their current state-of-the-art errorbars, which highlights the importance of accurate and precise determination of exoplanetary orbital parameters when attempting to perform morning-to-evening spectroscopy 
as we do in our work (see Methods).}

% We perform simple modelling around the H$_2$O ($2.84\ \mu$m) and CO$_2$ ($4.38\ \mu$m) features by approximating them as Gaussians in order to study whether they are consistent with each 
% other between the morning and evening terminator using a Bayesian evidence detection 
%framework similar to the one used in \cite{prism:2023}. We find that while the H$_2$O 
%features seem to be consistent between the morning and the evening spectra, the CO$_2$ 
%features do show differences at a greater than $\sim 3.6\sigma$ level.

Next, we performed forward modeling and atmospheric retrievals to explore the physical mechanism behind the difference in the observed morning and evening spectra. In particular, we used the ATMO forward model grid \cite{goyal:2018} and the chemically-consistent CHIMERA retrieval framework \cite{CHIMERA} with modifications that allow them to perform morning/evening inference simultaneously and account for the covariance between these limb depths \cite{EJ:2021}. We thus fitted the morning and evening spectra of WASP-39~b simultaneously to constrain their individual temperatures, C/O ratios, and cloud properties (see Methods for details). The best-fit solutions from both of those frameworks are qualitatively similar; we show the CHIMERA median model as solid blue and red lines for the morning and evening respectively in Figure \ref{fig:morning-evening-spectra}. Both modeling frameworks return similar constraints on the morning and evening temperatures and C/O ratios. Our CHIMERA retrievals converge to a $\textnormal{C/O}_\textnormal{evening} = 0.58^{+0.13}_{-0.16}$ and $\textnormal{C/O}_\textnormal{morning} = 0.57^{+0.17}_{-0.23}$, which are consistent with 
each other. These are, in turn, consistent with the ratios derived from the transmission spectroscopy analysis reported by the ERS team on the NIRSpec/PRISM observations \cite{prism:2023}, i.e., C/O $\sim 0.3-0.5$. Our derived metallicity (assumed to be common for both limbs) is also consistent with the $\sim 10$ times solar metallicity reported in that work. Interestingly, the CHIMERA retrievals support a significantly hotter, $1068^{+43}_{-55}$ K evening when compared to the morning retrieved temperature of $889^{+54}_{-65}$ K; the difference being $177^{+65}_{-57}$ K, significant at more than a $3\sigma$ level. This is the first time temperatures and C/O ratios have been able to be constrained from the morning and evening of an 
exoplanet. This result qualitatively follows predictions from 3D GCMs, on which hotter evening limbs arise due to superrotating equatorial jets on highly irradiated exoplanets such as 
WASP-39~b \cite{Kataria:2016, kempton:2017, powell:2019, Lee:2023, tsai:2023}. In addition, while the 
retrieval suggests a relatively constrained cloud-top location in the 
evening limb at about $\sim 1-10$ mbar, the cloud-top location in the 
morning limb is relatively unconstrained, allowing multiple possible 
configurations given this data. This is likely a consequence of the 
relatively flatter morning spectrum combined with the relatively large uncertainties on our NIRSpec/PRISM limb spectra, which allows for a wide 
range of possibilities for the morning cloud properties. While this suggests that most of the 
variations observed in the NIRSpec/PRISM morning/evening limb spectra 
could be attributed to temperature differences between the morning and 
the evening limbs, the treatment of clouds as grey opacity sources in 
the frameworks used to perform our inferences might be preventing us 
from further constraining the cloud properties in the morning and 
evening limbs of WASP-39~b. 

%%%%%%%  Start GCM section draft by Maria %%%%%%%%%%%%%%%%%%%%%%%

\begin{figure}
    \centering
    \includegraphics[width=\linewidth]{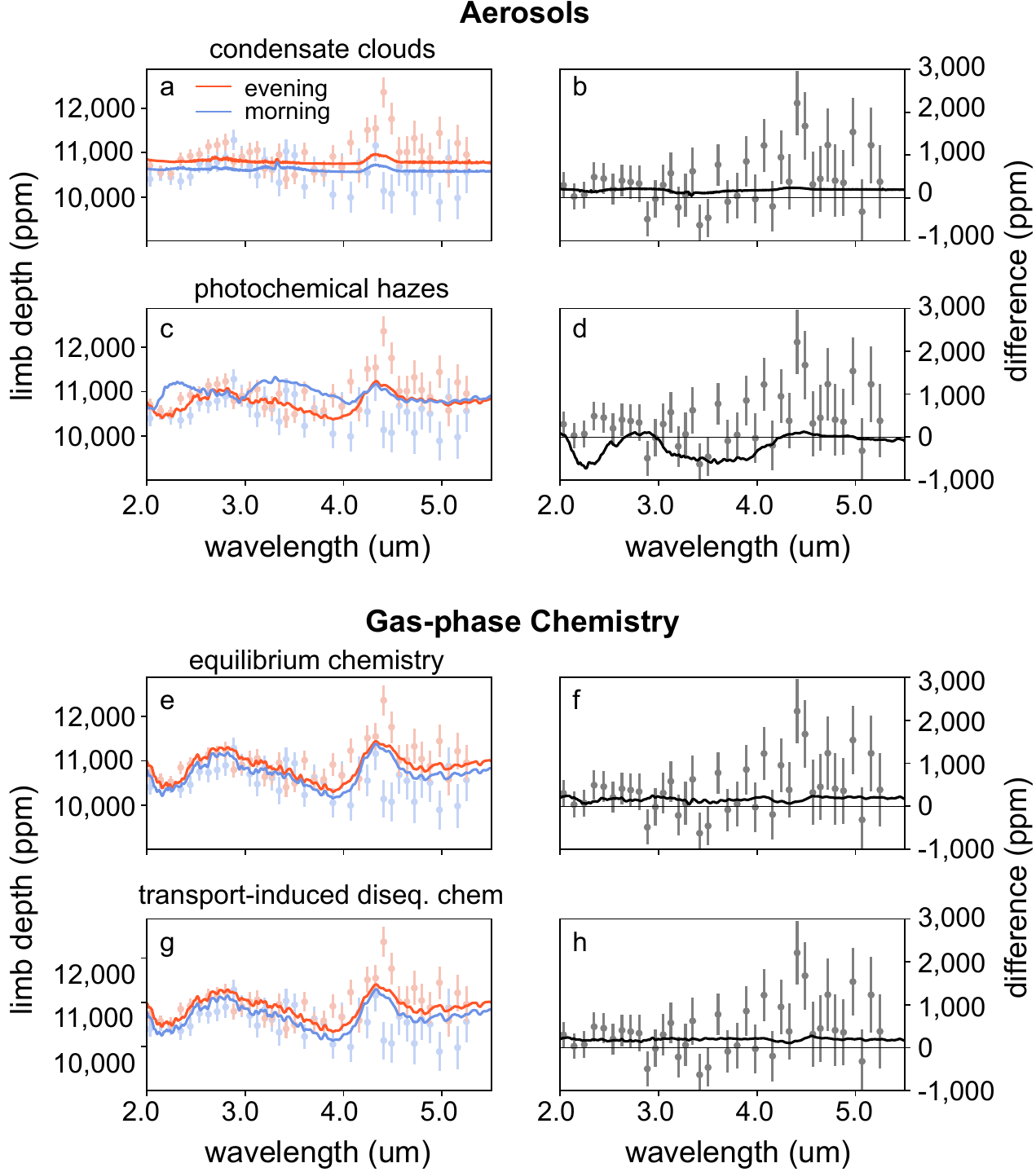}
    \caption{\textbf{Comparison of morning/evening spectra with General Circulation Models (GCM)s.} Comparison of predictions from general circulation models to the observed transit depths for the morning and evening terminator, as derived with \texttt{catwoman}. The left column shows the spectra, while the right column shows the difference between morning and evening terminator. {A vertical offset of -600~ppm was applied to the spectra in panels c, e, and g to facilitate comparison with the observed spectra.} \textbf{a-b.}  Condensate cloud model together with equilibrium chemistry taking into account elemental depletion by clouds. %postprocessed in 1D?
    {This model qualitatively provides the best match to the morning terminator spectrum.}
    \textbf{c-d.} Photochemical haze model based on SPARC/MITgcm, with equilibrium chemistry gas phase abundances.
    \textbf{e-f.} Clear-atmosphere equilibrium chemistry model.
    \textbf{g-h.} Clear-atmosphere model including transport-induced disequilibrium chemistry. {The clear-atmosphere models provide the best match to the evening terminator spectrum. All errorbars represent 1-standard deviation.}}
    \label{fig:gcms} 
\end{figure}

To explore the possibilities enabled by aerosols and chemical processes, we compared the observations to predictions from GCMs. GCMs are hydrodynamics models that simulate the 3-D wind and temperature structure in planetary atmospheres, self-consistently predicting differences between the evening and morning terminator. A range of processes influence terminator differences in the spectra, including temperature differences driven by atmospheric circulation, condensate clouds, photochemical hazes, and transport-induced disequilibrium chemistry of gaseous species (Fig. \ref{fig:gcms}).  No single GCM currently is able to self-consistently simulate all of these processes simultaneously. We thus included multiple different models to explore them \cite{CaroneEtAl2023WASP-39b, HellingEtAl2023ModelGrid, zamyatina:2023, SteinrueckEtAl2021}.

While the gas-phase chemistry results {(Unified Model)} reproduce well the shape of the spectral features {at the evening terminator}, the condensate cloud model {(ExpeRT/MITgcm)} performs qualitatively better at reproducing {the reduced feature amplitude at the morning terminator, especially the suppressed CO$_2$ feature at 4.3~$\mu$m. The photochemical haze model predicts large methane features at the morning limb and a larger morning limb in methane bands (centered at 2.3 and 3.3 $\mu$m), in conflict with observations. This is driven by colder temperatures at low pressures in this GCM (SPARC/MITgcm) combined with equilibrium chemistry. Photochemical hazes themselves have a smaller effect on limb differences. 
The condensate cloud model is the only one that predicts the largest limb difference to be in the center of the CO$_2$ feature (even though the difference is substantially smaller than observed). This finding is in line with another recently published cloud microphysics study \cite{arfaux2023}. A comparison with cloud-free spectra from the same GCM as our condensate cloud model (not shown) revealed that the limb difference in the center of 4.3 $\mu$m CO$_2$ feature is determined by the temperature difference but at other wavelengths is dominated by clouds. We thus suggest as tentative explanation a relatively cloud-free evening terminator and a cloudier morning terminator,} in line with the inferences performed with the ATMO forward models and CHIMERA retrievals described above. {Detailed follow-up studies exploring a larger parameter space with cloud formation models will be needed to test this explanation.}

\clearpage

\renewcommand{\figurename}{Extended Data Fig.}
\renewcommand{\tablename}{Extended Data Table}
\renewcommand{\theHfigure}{Extended Data Fig.~\arabic{figure}}
\renewcommand{\thetable}{Extended Data Table \arabic{table}}
\renewcommand{\theHtable}{Extended Data Table \arabic{table}}
\setcounter{figure}{0}
\setcounter{table}{0}

\section*{Methods}\label{methods}

\subsection*{Dataset}
In this work, we use the \textit{JWST} NIRSpec PRISM dataset obtained for WASP-39~b as part of The \textit{JWST} Transiting Exoplanet Community Director’s Discretionary ERS program \cite{stevensonERS, beanERS} (ERS 1366; PIs: N. M. Batalha, J. L. Bean, K. B. Stevenson) and which was already introduced in the work of \cite{prism:2023}. We selected this dataset for the morning/evening exploration of WASP-39~b as, among the four ERS datasets, it has the widest wavelength coverage and presented relatively minor challenges in the light curve data analysis --- which is how we obtain constraints on the morning and evening of the exoplanet in this work. 

\subsection*{Data analysis}
We use the $\texttt{FIREFLy}$ pipeline's reduction of the dataset as presented in \cite{Rustamkulov:2022}, \cite{prism:2023}, and subsequently used in Carter \& May et al. 
(in {review}). To summarize, the pipeline performs initial calibrations using the package $\texttt{jwst}$, adding in group-stage 1/$f$ noise destriping before the ramp-fitting stage. It then scrubs the time series of bad pixels and cosmic rays, and shift-stabilizes each integration to correct for millipixel-level correlated jitter of the spectral trace. Following Carter \& May et al. (in {review}), who delve deeply into the deleterious effects of saturation, we elect to use only the unsaturated 2.0-5.0 $\mu$m region of the spectrum due to its weak systematic noise and reproducible transmission spectrum. The $\sim$0.7 -- 2.0 $\mu$m region shows significant deviations relative to the unsaturated NIRISS-SOSS spectrum \cite{feinsteinERS}, increasing in magnitude toward the saturation center at 1.3 $\mu$m. The region also suffers from significantly lower signal-to-noise because fewer groups are available for use there. We omit this region to avoid the possibility of drawing spurious conclusions about the planet's nature. We note that WASP-39 b's 2.0-5.0 $\mu$m spectrum has relatively larger feature amplitudes spanning more chemical species than the NIR region. The PRISM spectrophotometry is well-fit by a linear trend varying in wavelength and displays no other significant systematic noise.

\subsection*{Transit light curve analysis}

\textbf{Physical and orbital parameters of the system.} The physical and 
orbital parameters of WASP-39~b used in this work are the ones reported by 
Carter \& May et al. (in review). These were fixed in our wavelength-dependent 
light curve fits. In particular, we fixed the period to $P = 4.0552842$ days, the 
scaled semi-major axis to $a/R_* = 11.390$, the impact parameter to $b = 0.4498$ 
and set the mid-transit time for NIRSpec/PRISM to $T_0 = 2459771.335647$ days.\\ 

\noindent \textbf{Wavelength-dependent light curve analysis.} In order to explore the evidence for morning/evening signatures on the wavelength-dependent NIRSpec/PRISM light curves, we decided to perform analyses following three different approaches: 

\begin{enumerate}
    \item \textit{\texttt{catwoman} light curve fits.} In this approach, we performed light curve fits to the wavelength-dependent light curves using the \texttt{catwoman} framework 
    introduced in \cite{EJ:2021, JE:2020}. This framework models the transiting 
    object passing in front of the star as two stacked semi-circles, each one representing the effective transit depth from the morning and evening 
    terminators of WASP-39~b. We performed two sets of fits using this approach, 
    led by co-authors MM and NE, respectively, both of which set the rotation angle of the limbs to 90 degrees. The NE \texttt{catwoman} fits 
    were performed by using limb-darkening coefficients taken from the \texttt{limb-darkening} \cite{EJ2015} package as priors, using ATLAS models with the same 
    stellar parameters as WASP-39, and passing those through the SPAM algorithm 
    of \cite{Howarth:2011} to obtain the final estimates for the limb-darkening 
    coefficients. The MM fits fixed the limb-darkening coefficients 
    to those obtained from the ExoCTK \cite{exoctk} library also obtained from 
    ATLAS models with the same stellar parameters as WASP-39. Both light curve fits 
    used a quadratic limb-darkening law. The NE fits used the limb-darkening parametrization in \cite{Kipping:2013}, with a prior that follows a truncated normal centered around the theoretical, transformed coefficients $(q_1, q_2)$, and with limits set between 0 and 1 for each. The standard deviation of that prior distribution was set to 0.1 for both transformed 
    coefficients based on the findings of \cite{PE:2022} which found $0.1$ to be 
    the maximum offset between quadratic and empirical limb-darkening coefficients 
    when comparing theoretical coefficients to those obtained via precise \textit{TESS} 
    photometry. Both fits also included a slope in time as a free 
    parameter to model the visit-long slope seen in the NIRSpec/PRISM data in \cite{prism:2023}, along with 
    a baseline flux offset. The NE fits include in addition a jitter term added in 
    quadrature to the NIRSpec/PRISM light curve errorbars. In total, the MM fits had 4 free parameters (2 semi-circle radii, slope and baseline flux offset) for each 
    wavelength-dependent lightcurve, whereas the NE fits had 7 free parameters 
    (same as MM plus 2 limb-darkening coefficients and a jitter term). \\
    
    \item \textit{Ingress/egress (\texttt{Tiberius}) light curve fits.} In this approach, we performed transit 
    light curve fits to the wavelength-dependent light curves using the 
    \texttt{batman} \cite{kreidberg2015batman} framework via the \texttt{Tiberius} library \cite{tib1, tib2}, but fitting 
    half-ingress (where contributions mainly from the morning limb are expected) 
    and half-egress (where contributions mainly from the evening limb are 
    expected) independently. To find the contact points of the transit event, a 
    transit lightcurve was generated with the orbital parameters described above, 
    and limb-darkening was set to zero to find the discontinuity points in the 
    light curve marking contact points 1 (start of ingress), 2 (end of ingress), 
    3 (beginning of egress) and 4 (end of egress). With this, the 
    half-ingress (contact point 1.5 --- mean of contact 1 and 2) and half-egress 
    (contact point 3.5 --- mean of contact 3 and 4) contact points were derived. 
    The half-ingress (all datapoints prior to contact point 1.5 and post contact point 4) and half-egress (all datapoints post 
    contact point 3.5 and prior to contact point 1) light curves were then fit as follows. First, the full lightcurve is 
    fit for each wavelength bin, with a model that fits for a slope and baseline flux 
    offset, the planet-to-star radius ratio and the linear term of a quadratic limb-darkening law --- with the quadratic term fixed to limb-darkening coefficients obtained from 3D models in the Exo-TiC library \cite{GW:2022}. Then, the half-ingress and half-egress lightcurves are fit by leaving all parameters fixed to 
    the best-fit parameters obtained from the full transit fit, with the only free 
    parameter being the planet-to-star radius ratios. These define the morning and 
    evening transit depth, respectively, using this methodology. \\
    
    % \item \textit{Wavelength-dependent mid-transit time.} As described 
    % by previous works \cite{VP:2016, powell:2019, EJ:2021}, the time-of-transit center might be degenerate with exoplanetary limb asymmetries. In this approach, we take advantage of this degeneracy by measuring the time-of-transit center as a function of wavelength, aiming to detect the impact of these limb asymmetries by studying this ``time" spectrum. This approach was already shown and described for WASP-39~b NIRSpec/PRISM data in \cite{prism:2023}. In summary, transit light curve fits were performed to the NIRSpec/PRISM wavelength-dependent light curves using the \texttt{batman} \cite{kreidberg2015batman} framework, with both the planet-to-star radius ratio \textit{and} the time-of-transit center left as free parameters. In addition, the limb-darkening coefficients, a slope, and a flux offset were also set as free parameters. 

    \item \textit{Wavelength-dependent mid-transit time.} {The transit time spectrum is sensitive to 0th-order wavelength-dependent opacity-centroid shifts along the planet's orbit imparted by spatial inhomogeneity in the composition and temperature of its terminator \cite{dd:2012, VP:2016, powell:2019, EJ:2021}. A hotter, more extended trailing terminator casts a positive deflection in the relative transit time ($\Delta T_0 > 0$), with the planet appearing to transit slightly later due to the subtle trailing-limb inflation of its $\tau=1$ surface. A colder leading terminator spectrum, or one with feature-muting grey clouds, would likewise cast a monochromatic positive deflection to the transit time spectrum. The effect amplitude scales inversely with the planet's orbital velocity and impact parameter, and the difference of each terminator's mean transit cord altitude. The time spectrum shown in Extended Data Figure \ref{fig:all-data}c is the result of a Levenburg-Marquardt least-squares fit to the spectrophotometry, with bins wider than those used in \cite{prism:2023}. At each wavelength channel we fix WASP-39 b's orbital parameters to those of Carter \& May et al. (in {review}), while fitting for its limb darkening coefficients, transit depth, and transit center time using \texttt{batman} \cite{kreidberg2015batman}. We find that the time signature is robust against one's selection of fitted and fixed parameters, resulting in $\leq 1-\sigma$ differences. Statistically-significant $\geq 3-\sigma$ positive deflections are detected with wavelengths corresponding to the spectral features of H$_2$O ($\sim$ 2-3.5 $\mu$m), H$_2$S ($\sim$ 3.78 $\mu$m), SO$_2$ ($\sim$ 4.06 $\mu$m), and CO$_2$ ($\sim$ 4.2-4.5 $\mu$m)}. 
    
\end{enumerate}

\begin{figure}
    \centering
    \includegraphics[width=\linewidth]{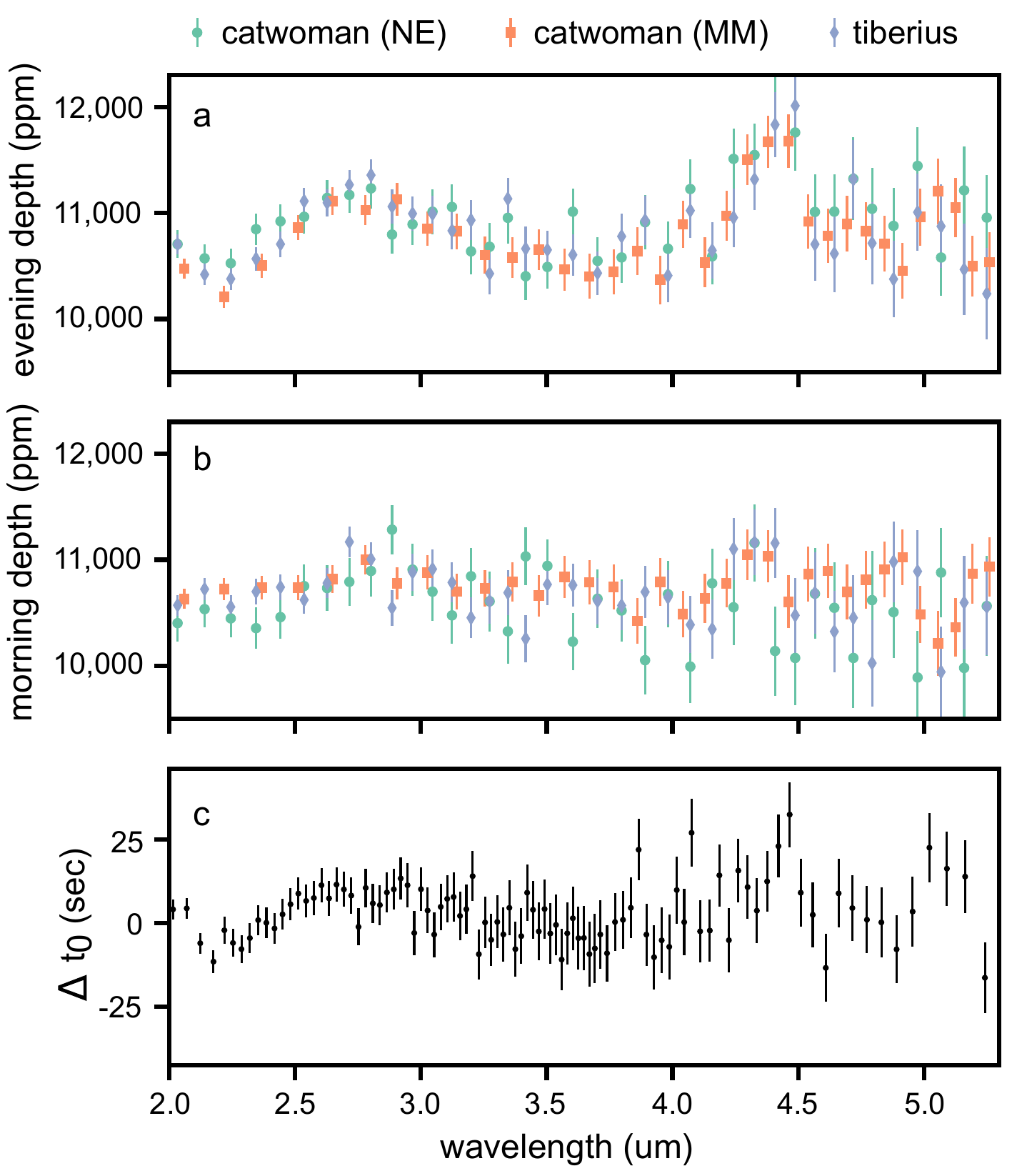}
    \caption{\textbf{Different approaches at detecting limb asymmetries from NIRSpec/PRISM data.} \textbf{a-b.} Evening (top) and morning (middle) depths as extracted from three independent analyses of our NIRSpec/PRISM lightcurves; one using the \texttt{catwoman} framework with limb-darkening as free parameters with a prior (NE), a framework leaving those fixed (MM) and a framework on which half-ingress and half-egress are fitted independently using a \texttt{batman} lightcurve model (Tiberius); see text for details. Note the agreement between approaches for both terminators, and how the amplitude of the features seem to be smaller in the morning terminator \textbf{c.} An independent look at limb asymmetries by fitting for a wavelength-dependent time-of-transit center to each wavelength-dependent lightcurve. As with the top and middle panels, differences between the limbs as tracked by the time-of-transit center seem to be largest between 
    2-3.5 $\mu$m, i.e., around the water bands. {All errorbars represent 1-standard deviation.}}
    \label{fig:all-data} 
\end{figure}

The results from the methodologies described above are presented in 
Extended Data Figure \ref{fig:all-data}. Both \texttt{catwoman} (NE, MM) and the 
half-ingress/egress approaches seem to be consistent with each other within the errorbars, as well as judging from the overall structure of the resulting morning and evening spectra. In general, it appears the morning spectral features 
of H$_2$O and CO$_2$ are damped in comparison to the same features 
observed in the evening, a picture that is consistent with the analysis 
performed on the wavelength-dependent transit times, where the largest 
offsets are observed for the H$_2$O and CO$_2$ features, although the latter molecule is seen at lower significance. 

However, the significance of the 
dampening between spectral features in the morning and evening is less clear for the NE light curve analysis. This latter approach produces 
the largest errors on the morning/evening spectra, which is due to the 
fact that this approach assumes some ignorance on the limb-darkening 
coefficients. As we show below through light curve experiments performed 
to study the robustness of our morning/evening spectra to assumptions on 
transit parameters, we deem this approach as the most conservative in terms 
of deriving differences between the morning and evening limbs, as very 
slight deviations from the true, ``underlying'' limb-darkening coefficients 
can give rise to false differences between morning and evening spectra, 
which could in turn be erroneously interpreted as an astrophysical effect in 
the exoplanet atmosphere. This is the reason why we decide to perform inferences on the \texttt{catwoman} approach of NE to showcase/interpret 
morning/evening spectra in the main text.\\

\noindent \textbf{Robustness of morning/evening spectra to assumptions on transit parameters.} Previous works that have studied the possibility of extracting morning and evening spectra directly from transit light curves  have identified a number of possible degeneracies that might arise 
and which could impact the derived spectra \cite{VP:2016, powell:2019, EJ:2021}. In particular, the degeneracy between 
the time-of-transit center and limb asymmetries has been identified in previous works as the largest source preventing performing these 
detections in real data, with the usage of precise wavelength-dependent transit 
light curves like the ones used in this work being critical to lift it \cite{VP:2016, powell:2019}. In addition, while limb-darkening has been shown 
to slightly decrease the detectability of limb asymmetries on simulated data 
\cite{EJ:2021}, any biases induced by fixing those coefficients have, to our 
knowledge, not been studied in detail in the literature. Finally, eccentric orbits have been shown to also give rise to asymmetric transit light curves \cite{Barnes:2007, Kipping:2008}. Using the wrong assumption about the eccentricity of the orbit could, thus, in turn, give rise to biases on morning/evening spectra 
which are extracted by measuring light curve asymmetries. We performed experiments to study those three systematic error sources in order to 
quantify their impact on our reported morning/evening spectra.\\

\noindent \textit{Robustness against no limb asymmetries.} In order to define 
a null baseline robustness check against the \texttt{catwoman} methodology 
used in this work, we simulated noisy transit light curves using 
\texttt{batman} which had the same input transit parameters as the 
white-light light curve parameters used in our fits (i.e., the ones from 
Carter \& May et al., in review), the same noise properties as the real 
data (including the visit-long slope), and transit depths 
that varied as a function of wavelength matching the transit spectrum 
presented in \cite{prism:2023}. We then fitted those lightcurves with a \texttt{catwoman} model, with the same approach as the one described 
above for the NE \texttt{catwoman} light curve fits. As expected, we 
found that the extracted morning 
and evening spectra were consistent with each other, producing a null 
difference between them. The results of those simulations are shown in 
Extended Data Figure \ref{fig:experiments}, panel \textbf{a.}\\

\noindent \textit{Robustness against the time-of-transit center.} In order to check the robustness of our spectra to using a fixed time-of-transit 
center (obtained from the work of Carter \& May et al., in review), 
we performed the same simulations as the ones described in the previous paragraph; however, we used a time-of-transit center being $3\sigma$ larger than the one reported in the work of Carter \& May et al. (in review), which amounted to a time offset of 3.4 seconds. We then fitted those 
light curves with a \texttt{catwoman} model which used the time-of-transit 
center as a fixed parameter without this offset. We found that such an offset in 
timing had no measurable impacts on our morning/evening transit spectrum, with the difference between them being consistent with zero. The results of those simulations are shown in 
Extended Data Figure \ref{fig:experiments}, panel \textbf{b.}\\

\noindent \textit{Robustness against limb-darkening coefficients.} We performed a similar experiment to the one described in the previous paragraph but modified the input limb-darkening coefficients to be offset by 0.01 from 
the fixed values, which were obtained following the methodology 
described above for NE's \texttt{catwoman} transit light curve fitting 
approach.  We then fitted those light curves, but fixed the limb-darkening 
coefficients in our fit to the ones without those offsets. 
We found that these very small offsets on the limb-darkening coefficients 
had a measurable impact on the retrieved morning/evening spectra, giving 
rise to measurable morning/evening differences of order $\sim 200-300$ 
ppm when fixing those in the light curve fitting procedure to the wrong values; 
the results from this simulation are presented in Extended Data Figure 
\ref{fig:experiments}, panel \textbf{c.}. The same experiment, but 
setting wide priors on these coefficients as described on the NE 
\texttt{catwoman} light curve fitting procedure described above allows 
one to recover the input null difference between the evening and morning 
spectra (not shown). These experiments highlight that, while \textit{relative} morning/evening spectral differences 
might be obtained by fixing the limb-darkening coefficients even if they are 
slightly wrong, \textit{absolute} spectral differences might not be robustly 
extracted in general. Even if aiming to obtain relative morning/evening spectral differences, in reality limb-darkening offsets might be wavelength-dependent and thus might give rise to spurious signals and/or spectral features in the limb spectra. This was one of the main reasons 
why we decided to present the results obtained from the NE 
\texttt{catwoman} light curve fitting procedure in the main text, which 
allows for limb-darkening coefficients to be significantly offset from the 
input theoretical model calculations. The retrieved limb-darkening 
coefficients when performing \texttt{catwoman} fits using the NE approach 
to our real NIRSpec/PRISM data are presented in Extended Data Figure \ref{fig:ldcs}. 
Errors on the $u_1$ coefficients range from $0.02-0.04$, while errors on the $u_2$ 
coefficients range from $0.03-0.06$. Offsets on the limb-darkening coefficients of 
order $\sim 0.01$, thus, are indeed allowable by the data, and they are particularly 
likely in the 2.5-4.5 $\mu$m range for the linear ($u_1$) coefficient of the 
quadratic law which is where the retrieved limb-darkening coefficients deviate the 
most from the theoretical model predictions.\\

\noindent \textit{Robustness against eccentricity.} We repeated a similar 
experiment to the ones described above, but this time we set as an input 
\texttt{batman} model an eccentric orbit with parameters consistent at 
$3\sigma$ with the best-fit parameters presented in Carter \& May et al. (in review), 
which corresponds to $e=0.035$ and $\omega = 10$ degrees. Our simulations, 
presented in Extended Data Figure \ref{fig:experiments}, panel \textbf{d.}, show that 
these set of values can indeed give rise to offsets between the morning and evening 
limbs, making the morning limb spectra larger than the evening 
limb spectra. We explored the range of ($e$, $\omega$) allowed by the analysis 
presented in Carter \& May et al. (in review) and found that the impact of these 
parameters act always in the same direction for the WASP-39~b transit observations 
analyzed in this work: values of $e$ and $\omega$ allowed by the posterior distribution 
all could give rise to larger morning depths than evening depths. With the 
reported morning/evening spectra in this work we observe the opposite, however: 
larger evening depths than morning depths. This suggests that the absolute difference 
we observe between the morning and evening spectra in the NIRSpec/PRISM observations 
are, at worst, lower limits on the actual absolute depth difference between the morning and 
the evening limbs. 

It is important to note, however, that a set of values of 
($e$, $\omega$) as the one used for our experiment is likely unrealistically large, 
as secondary eclipse observations constrain $e\cos \omega = 0.0007\pm 0.0017$ \cite{Kammer:2015}, which would reject such ($e$, $\omega$) combination at more than 
10-$\sigma$. Given WASP-39~b's eccentricity is consistent with zero given data from different sources, which is in turn consistent with the relatively small circularization 
time-scales for the planet given WASP-39 is a $\sim 9$ Gyr star \cite{Faedi:2011}, 
we suggest eccentricity offsets to be a relatively minor effect for this system. 

{To test how the most up-to-date constraints on the properties of the 
system impact our detection of morning and evening differences in the transmission 
spectrum of WASP-39~b presented in this work, we decided to re-run a white-light lightcurve 
fit similar to that in Carter \& May et al. (in review) but assuming an eccentric orbit, 
and then let the \textit{posteriors} of the transit parameters of this fit float as priors 
on our wavelength-dependant lightcurve fits instead of fixing those parameters as was 
done in our nominal analysis. This new white-light lightcurve fit was performed 
with two extra priors: (1) the constraint on $e\cos \omega = 0.0007\pm 0.0017$ of \cite{Kammer:2015} and (2) a prior 
on the stellar density of WASP-39~b obtained via the methodology outlined in \cite{brahm1} and \cite{brahm2}. This latter methodology first obtains stellar atmospheric parameters obtained 
from high-resolution spectra, for which we use the average of 3 
publicly available FEROS high-resolution spectra from PID 098.A-9007(A) obtained in 
February 12, 2017, which were reduced with the CERES pipeline \cite{ceres}. This spectrum 
is given as input to the ZASPE code \cite{zaspe}, with which an initial set of stellar 
atmospheric parameters is obtained. Then, the available Gaia 
($G=11.8867 \pm 0.0020$, $B_{P}= 12.3061 \pm 0.0054$, $R_{P} = 11.3258\pm 0.0031$) and 
2MASS ($J = 10.663 \pm 0.024$, $H = 10.307 \pm 0.023$, $K = 10.202 \pm 0.023$) photometry is 
combined with Gaia-derived distances ($211.46\pm2.35$ pc; obtained via parallaxes using the methodology described in \cite{BJ}) to obtain fundamental, absolute stellar parameters using PARSEC 
isochrones \cite{parsec}, with the spectroscopically derived stellar atmospheric parameters used as priors. This iterative procedure returns all the fundamental stellar parameters for the star. 
In particular, for WASP-39~b we find $R_* = 0.897 \pm 0.011 R_\odot$, $M_* = 0.891 \pm 0.033 M_\odot$, 
which in turn gives a stellar density of $\rho_* = 1736 \pm 121$ kg/m$^3$. 
We note those estimated stellar parameters are consistent with, albeit more precise than, 
those presented in the discovery paper of \cite{Faedi:2011}.}

\begin{figure}
    \centering
    \includegraphics[width=1.0\linewidth]{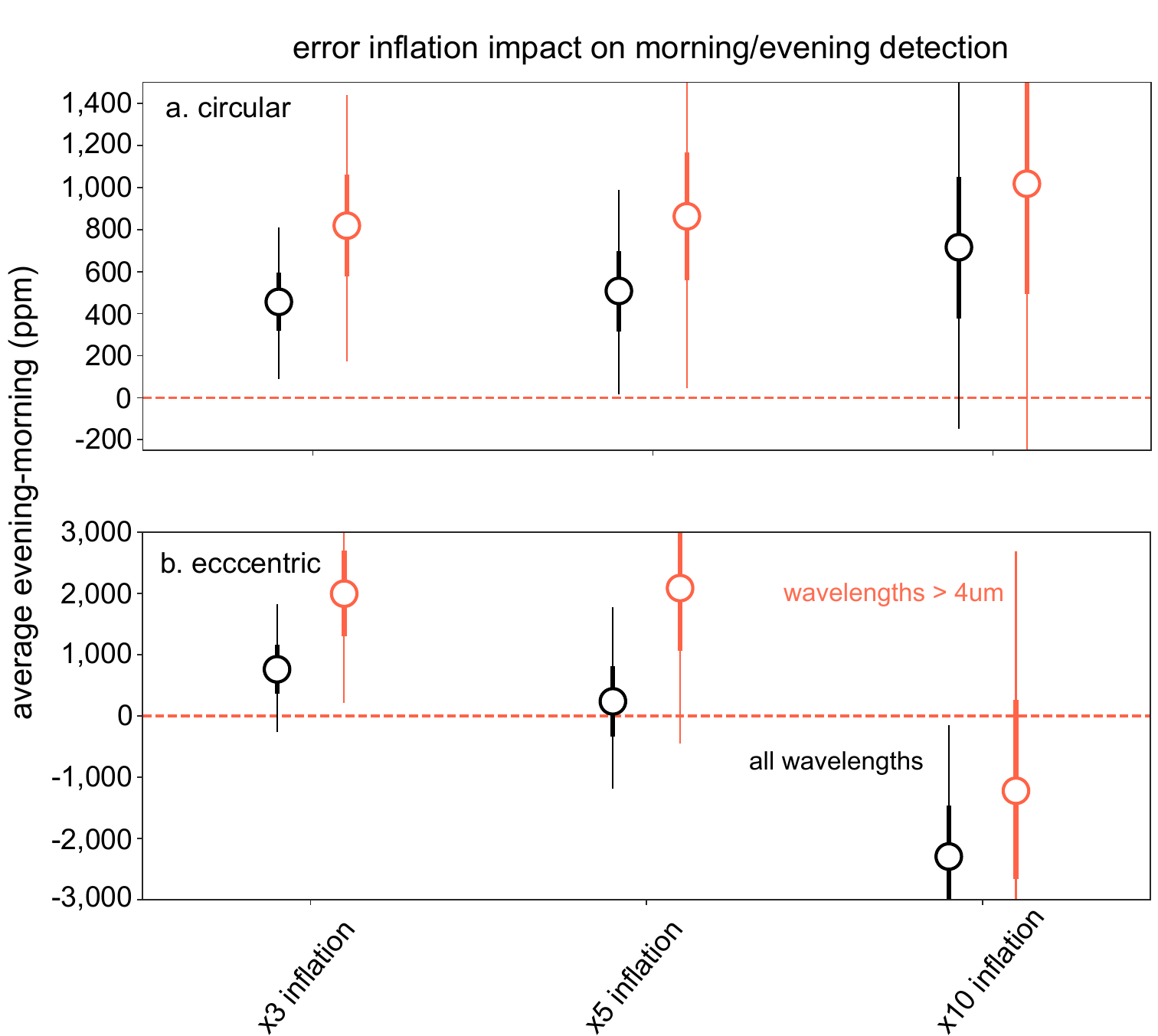}
    \caption{{\textbf{Impact of less accurate and precise orbital parameters on the detection of inhomogeneous terminators on WASP-39~b.} In our experiments, we inflated the errorbars on the orbital parameters (e.g., impact parameter, $a/R_*$, etc.) by different factors, and performed wavelength-dependant \texttt{catwoman} lightcurve fits on our NIRSpec/PRISM data using normal priors for each parameter along with the other wavelength-dependant parameters described in the main text and in the Methods section such as the planet-to-star radius ratio, limb-darkening, etc. \textbf{a.} Error inflation exercise assuming a circular 
    orbit (i.e., eccentricity fixed to zero) and \textbf{b.} same exercise but 
    assuming an eccentric orbit --- with uncertainties on all parameters, including 
    $e\cos \omega$ and $e\sin \omega$, inflated by 3 (left), 5 (middle) and 10-fold (right). Dashed line marks the non-detection threshold (i.e., equal evening 
    and morning depths). All bold errorbars represent 1-standard deviation. Thin errorbars are 3-standard deviations.}}
    \label{fig:errorinflation} 
\end{figure}

\begin{figure*}[t]
    \centering
    \includegraphics[width=0.496\textwidth]{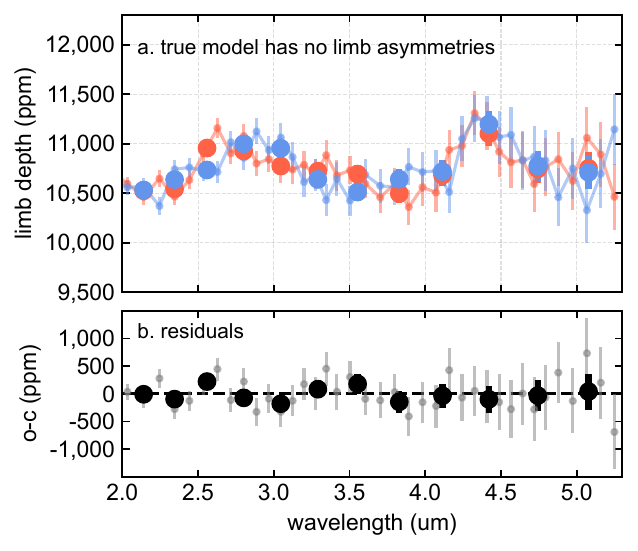}
    \includegraphics[width=0.496\textwidth]{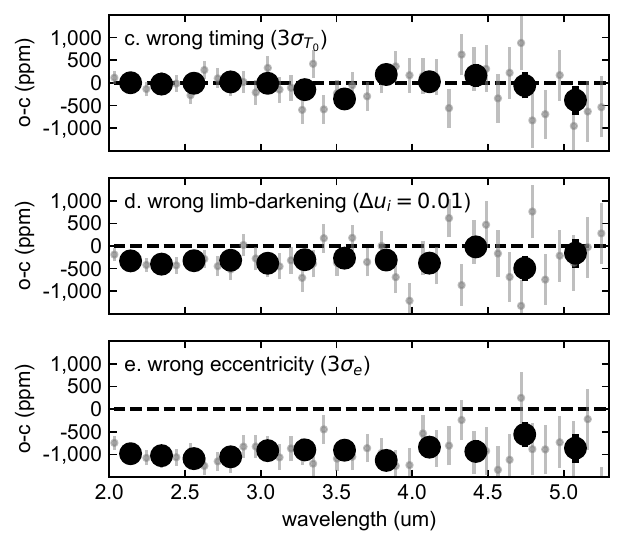}
    \caption{\textbf{Robustness of limb asymmetry detection to transit parameters and assumptions.} To study the robustness of our extracted morning and evening spectra for the NIRSpec/PRISM 
    observations, we simulated transit lightcurves using 
    \texttt{batman} and then fitted those using \texttt{catwoman} with different assumptions, leaving all parameters fixed but the depths of 
    the morning and evening limbs. \textbf{a-b.} Null case on which the 
    true model has no limb asymmetries and the input transit parameters are unchanged; the \texttt{catwoman} fits 
    correctly recover the same morning (blue) and evening (red) spectra 
    (top). The difference $\Delta$ between the morning and the evening 
    spectra are consistent with zero, as expected from this null case (bottom). \textbf{c.} Same experiment 
    as in a., but generating a light curve with a time-of-transit center 3-$\sigma$ away from the fixed value, which amounts to an offset of 3.4 
    seconds. The difference $\Delta$ is consistent with zero, suggesting 
    our inferences are robust against this parameter. \textbf{d.} Same 
    experiment, but generating a light curve that had limb-darkening 
    coefficients of the quadratic law offset by 0.01. Note how this injects 
    a systematic offset in the difference between the morning and evening 
    spectra; depending on the direction of this offset, this can lead to 
    mornings having larger depths than evenings or viceversa. \textbf{e.} 
    Same experiment, but generating a transit lightcurve with a non-zero 
    eccentricity consistent at 3-$\sigma$ with the white-lightcurve fits of Carter \& May et al. (in review; $e=0.035$, $\omega = 10$ deg). Note how this slight eccentricity can 
    generate significantly larger mornings than evenings, due to the asymmetry an eccentric orbit imprints on the transit lightcurve. For WASP-39~b, this eccentricity effect cannot generate larger evenings than mornings, which is what we observe. This suggests our results are also robust against this parameter (see text for details). {All errorbars represent 1-standard deviation.}}
    \label{fig:experiments} 
\end{figure*}

\begin{figure}[t]
    \centering
    \includegraphics[width=\linewidth]{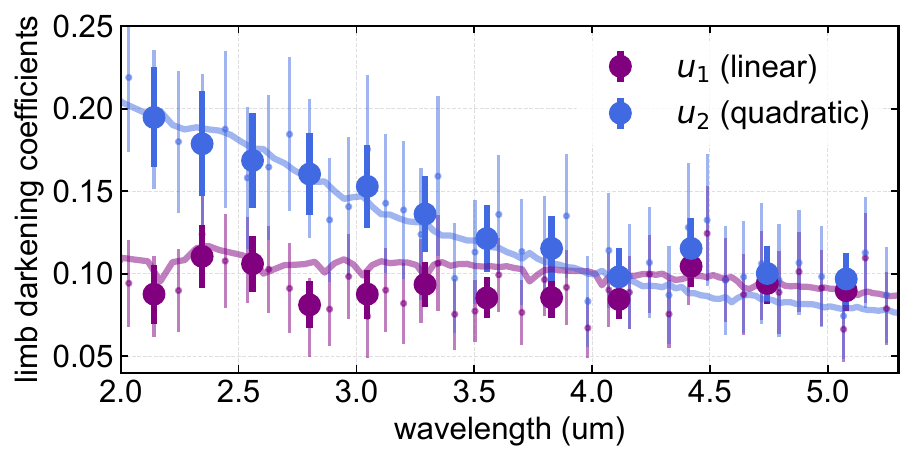}
    \caption{\textbf{Quadratic limb-darkening coefficients from \texttt{catwoman} WASP-39~b transit light curve fits.} Solid lines (purple for $u_1$, blue for $u_2$) are the 
    theoretical limb-darkening coefficients obtained by, first, using 
    the \texttt{limb-darkening} library using ATLAS models to extract 
    limb-darkening coefficients, and then passing those through the SPAM 
    algorithm of \cite{Howarth:2011} to obtain the model predictions. 
    Points with errorbars are retrieved limb-darkening coefficients from 
    our \texttt{catwoman} (NE) transit light curve fits. Note the apparent 
    offset between model and retrieved coefficients between about 2.5 to 4.5 $\mu$m for $u_1$. {All errorbars represent 1-standard deviation.}}
    \label{fig:ldcs} 
\end{figure}

{This new white-light lightcurve fit gives posterior parameters that are, in turn, all consistent with the 
ones reported in Carter \& May et al. (in review); notably, with a stellar density posterior 
of $1705 \pm 28$ kg/m$^3$, consistent at 1-sigma with the value of $1699 \pm 55$ kg/m$^3$ shown in that work (and much more constrained than our stellar density prior, showcasing this value to 
not be dominated by it). The fit also puts a limit on the eccentricity of WASP-39~b of $e<0.016$ with 99\% credibility ($e\sin (\omega) = 0.0014^{+0.0051}_{-0.0057}$; $e\cos (\omega) = 0.0009^{+0.0014}_{-0.0012}$); a circular model still being preferred via the 
bayesian evidence. As described above, we then use the posterior distributions of all the transit parameters of this eccentric model fit as priors for each wavelength-dependant lightcurve fit using the \texttt{catwoman} NE methodology described above. We find a morning and evening transmission spectrum which is 
very similar to the one presented in Figure \ref{fig:morning-evening-spectra}, albeit with 
larger errorbars; an increase dominated by the uncertainty on the eccentricity. This highlights the importance of having high-precision orbital parameters for systems when performing limb asymmetries detections, and in particular good constraints on eccentricities and arguments of periastron. Despite 
these enlarged errorbars, we do still find an average evening-to-morning depth difference 
of $582\pm 188$ ppm --- a 3-sigma detection of the evening-to-morning depth effect 
presented in this work even in this worst-case scenario. {In order to further explore the impact of our orbital parameter estimation on our detection of inhomogeneous terminators on WASP-39~b, we ran experiments on which we artificially enlarged the errorbars on the orbital parameters by 3, 5 and 10-fold, for both circular (i.e., $e=0$) and eccentric (i.e., free $e\cos (\omega)$ and $e\sin (\omega)$) cases, and compared the average evening-to-morning depths for each case. We compared these differences both across the entire wavelength range and for wavelengths above 4 $\mu m$, where we see the largest deviations on evening-to-morning depths. The results from our experiment are presented in Extended Data Figure \ref{fig:errorinflation}. As can be observed, for our circular case, error inflations of up to 5 times our reported errorbars still support our detection of inhomogeneous terminators, whereas for the eccentric case, error inflations of up to 3 times only support this detection at wavelengths above 4 $\mu m$. These experiments highlight the dependency on the detection of this effect on the accuracy and precision of orbital parameters for WASP-39~b --- which might also be the case for detecting the effect on other exoplanets. We do note, however, that those experiments} provide an upper 
limit on the uncertainty on the evening-to-morning depth difference, as the eccentricity 
should be wavelength-independent. In our case, however, we are fitting for eccentricity 
independently on each individual wavelength bin.} \\

\noindent \textit{Robustness against white-light limb asymmetries.} The 
work of Carter \& May et al. (in review) performed white-light light curve 
analyses assuming a \texttt{batman} transit light curve model. We leave the transit 
parameters fixed in most of our wavelength-dependent light curve fits to the 
NIRSpec/PRISM data. However, given our detection of limb asymmetries on this dataset, it 
might be suggested that the posterior parameters in Carter \& May et al. 
(in review) could be biased given no \texttt{catwoman} models were used to analyze 
the \textit{JWST} data. We performed such fits leaving all the priors for the rest of the 
parameters unchanged and as described in Carter \& May et al. 
(in review), but allowing for the \textit{JWST} data to have asymmetric limbs through 
a \texttt{catwoman} model. We found all the transit parameters agreed within 
$1\sigma$ with respect to the values reported in Carter \& May et al. (in review); 
however, the uncertainties in the time-of-transit center, eccentricity and stellar 
density are larger by a factor of $\sim 1.5$, $\sim 2$ and $\sim 4$ for the \texttt{catwoman} fits. When accounting for the constraints imposed by WASP-39~b's 
secondary eclipse described in the previous paragraph, however, all uncertainties but 
that of the stellar density are consistent between the analyses. This analysis 
suggests that the fact the work of Carter \& May et al. (in review) did not use 
\texttt{catwoman} models to fit the \textit{JWST} transit light curves is not a 
particularly important consideration for the case of WASP-39~b, in particular when 
it refers to the constraints on the eccentricity and time-of-transit center used 
in our wavelength-dependent fits.\\

\noindent \textbf{Robustness of morning/evening spectra to planetary rotation.} The framework and results presented above omit any impacts from planetary rotation 
during transit. If this were to be significant, the spectra observed at the very beginning of the transit event could be different from the spectra that is observed at the end of it. Assuming the exoplanet is tidally locked, the amount of rotation the planet undergoes (period of $\sim 4.1$ days) on timescales of the transit event ($\sim 2.8$ hours) is of order $\sim 10$ degrees; calculations by \cite{Wardenier:2022} suggest this might be too small to detect any signatures due to planetary rotation. To explore these constraints on our data, we decided to study it by performing the same lightcurve analysis outlined above, but considering a different set of morning/evening depths during ingress than during egress. We found that during ingress, the mean morning-to-evening transit depth difference was $-228 \pm 187$ ppm; during egress, the difference was of $344 \pm 189$ ppm. Both are consistent with zero at $2\sigma$ and thus we are unable to detect any differences between the morning/evening spectra during ingress and egress. We then joined those morning/evening depths observed during ingress and egress into a single, 
total transit depth. When comparing the total transit depth during ingress to 
that of egress, we find a mean difference of $234 \pm 144$ ppm{. This suggests that, 
while it is possible rotation effects are indeed important, they remain hard to detect 
with the quality of data at hand}.\\

\noindent \textbf{Robustness of morning/evening spectra to stellar rotation.} Stellar rotation could, in principle, produce asymmetries in the transit lightcurves as the planet transits red and blueshifted regions of the star. WASP-39~b, however, has a very slow stellar 
rotation of $1.5\pm 0.6$ km/s \cite{Faedi:2011}; in addition, \textit{JWST}'s NIRSpec/PRISM's resolution of about $R=100$ suggests such an effect should be small in the case of our observations. We performed calculations to determine how big such an effect would be on our observations and concluded this is below 1 ppm even in the worst-case scenario of a stellar rotation speed at the 5-sigma limit imposed by the work of \cite{Faedi:2011}, i.e., $4.5$ km/s.\\ 

\noindent \textbf{Robustness of morning/evening spectra to stellar heterogeneities.} Stellar heterogeneities (caused by, e.g., spots and faculae) 
could, in principle, impact our ability to retrieve limb asymmetries from transit light curves. WASP-39, however, is a relatively quiet G-type star \cite{Faedi:2011}. While photometric variability has been detected 
in \textit{TESS} and \textit{NGTS} light curves with a low amplitude level of 0.06\% in \cite{ahrer:2023} --- on the lower tail of photometric variability observed in \textit{Kepler} for G8-type stars 
like WASP-39 \cite{TLE2} --- no evidence of spot-crossing events on transit 
light curves of WASP-39 has been detected to date. 

On the one hand, unnoculted stellar heterogeneities 
such as the ones modeled by the transit light source effect \cite{TLE2} would impact the limb spectra of the morning and evening in similar ways, and would thus be unable to give rise to the morning/evening differences observed in this work. Occulted cool or hot spots could, in principle, cause asymmetries in the transit light curve. These would have to be, however, larger than 
about $100$ ppm in amplitude in the transit light curve at about 4 $\mu$m 
to give rise to the $\sim$ 400 ppm differences we observe between morning and evenings (see Figure 1d). In turn, these should \textit{increase} in amplitude 
for shorter wavelengths, which does not match the wavelength dependance 
of the limb asymmetries observed in our work. In addition, any such features 
would be several times larger at optical wavelengths; however, although such 
features should be easily detectable, no such features where reported in 
the optical NIRSpec/PRISM light curves analyzed in \cite{prism:2023}.

Based on this, we suggest stellar activity is unlikely to give rise to the morning/evening differences 
observed in this work. 

%\subsection*{Model evidence \& Interpretation}

% Results for prior on limb-darkening coefficients:
%\begin{tabular}{lccc}
%\hline
%Morning/Evening model difference & N$_p$ & $\ln \mathcal{B}_{\textnormal{CO2}}$ & $\ln \mathcal{B}_{\textnormal{H2O}}$\\
%\hline
%\hline
%Baseline \& abundance difference & 4 & \textbf{-0.76} & 2.79 \\
%Baseline difference & 3 & \textbf{0} & \textbf{-1.29} \\
%Abundance difference & 3 & -2.0 & \textbf{0.96} \\
%No difference & 2 & -5.6 & \textbf{0}\\
%\end{tabular}

\subsection*{Forward models \& retrievals}

\textbf{ATMO model grid.} We fitted the entire ATMO local condensation grid tailored to WASP-39~b as introduced in \cite{goyal:2018}. These models include all the dominant species observed in the NIRSpec/PRISM spectrum of WASP-39~b, except for sulfur species, which are nonetheless marginally detected in this dataset \cite{prism:2023}. To fit the morning and evening spectra, we take into consideration the fact that both are highly 
correlated, and thus use the log-likelihood framework introduced in \cite{EJ:2021}. To construct the morning/evening depth models, we simply take half the transit depth of a 
given ATMO model sampled for either the morning or evening limb spectra. Then, we try all the combinations of models in the grid to fit the morning and evening limb depths. Given there are 3,920 individual models for WASP-39~b in this grid, this resulted in over 15 million fits.

To bring those models to the observed transit depths in our data, we anchored the 
mean transit depth of both models to the average evening limb transit depth. This 
conserves any morning-to-evening transit depth offsets; this offset is, thus, the only 
free parameter in our model fits. Our set of best-fit models are all consistent with 
models on which the evening spectrum is hotter by about $200$ K than the morning spectrum, 
spanning a wide range of possible cloud and haze properties. In terms of C/O ratios 
and metallicities, our best-fit models all produce similar C/O ratios for the morning 
and evening spectra in the range $0.3-0.6$, and metallicities on the order of $\sim 10$ times solar.\\

\noindent \textbf{CHIMERA atmospheric retrievals.} In order to perform a posterior exploration of the parameter space allowed by our observed morning and evening spectra, 
we decided to run atmospheric retrievals using the \texttt{CHIMERA} 
retrieval framework described in \cite{CHIMERA} and modified in \cite{EJ:2021} to 
handle morning and evening spectra. These models include all the dominant species observed in the NIRSpec/PRISM spectrum of WASP-39~b, except for sulfur species, which are nonetheless marginally detected in this dataset \cite{prism:2023}. This framework performs chemically-consistent modelling, i.e., performing chemical equilibrium calculations given C/O ratios, metallicites and temperature/pressure profiles, which are combined with a prescription for clouds following the work of \cite{AM:2001}. We used the same prior distributions introduced in 
\cite{EJ:2021}, with the only modification being the prior on the temperature of the 
limbs, which here we set to be a uniform prior between 500 and 2000 K for both the 
morning and the evening limb. Our atmospheric retrievals considered a common 
metallicity, fiducial 10-bar radius and parameters defining the temperature/pressure 
profile for the morning and the evening limbs, but considered different C/O ratios, 
vertical mixing and cloud-top properties for the morning and the evening limb.

\begin{figure*}
    \centering
    \includegraphics[width=\textwidth]{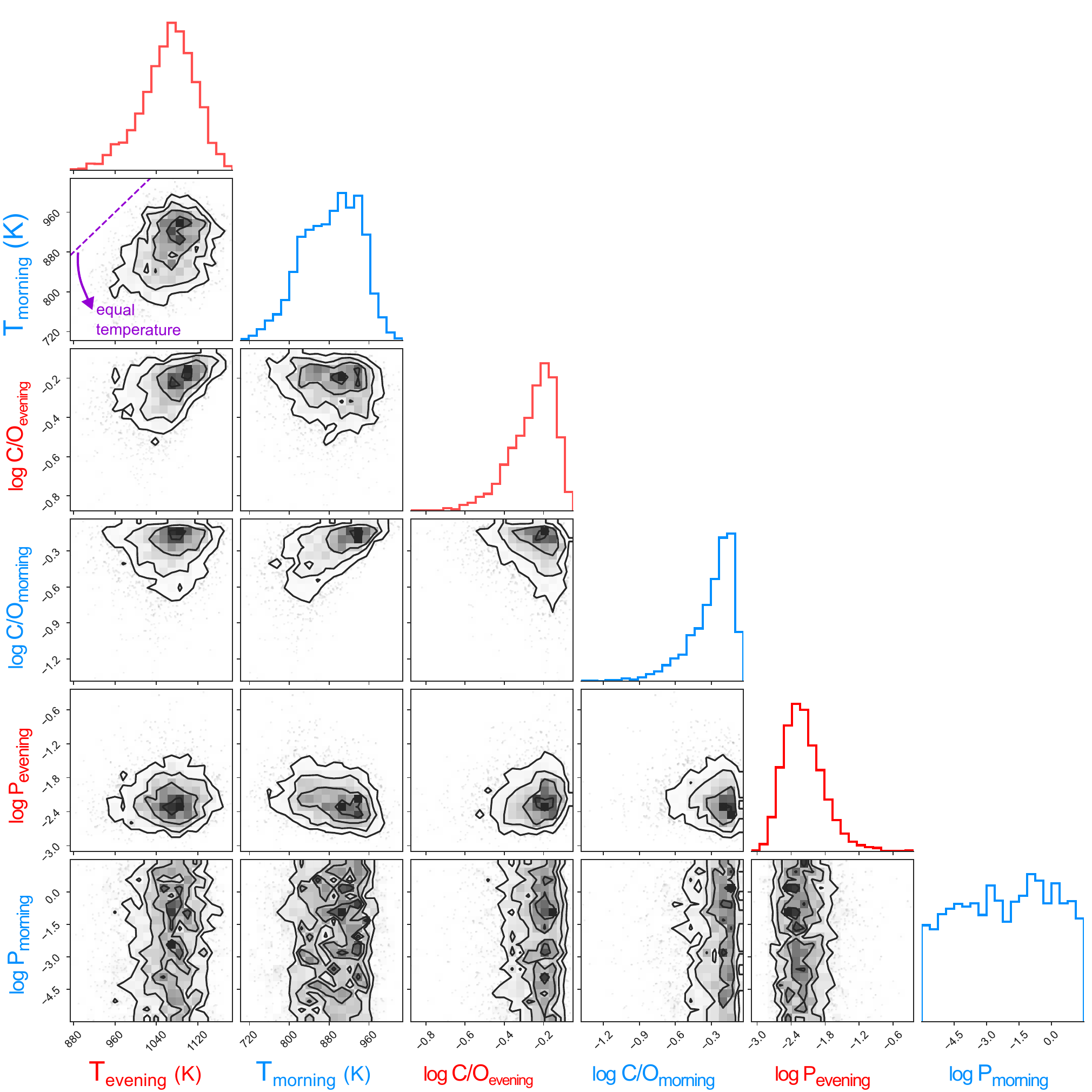}
    \caption{{\textbf{Posterior distribution of some of the retrieved CHIMERA parameters.} Corner plot of the morning/evening temperatures, (log) C/O ratios and (log) cloud-top pressures. Only the evening cloud-top pressure is constrained by our retrievals. The purple line in the morning/evening temperature posterior 
    samples showcases the line of equal temperatures; as can be observed, our posterior samples imply a significantly different morning-to-evening temperature.}}
    \label{fig:corner_plot} 
\end{figure*}

The posterior distributions from our \texttt{CHIMERA} atmospheric retrievals{ ---for which a set of posterior parameters are shown in Extended Data Figure \ref{fig:corner_plot} ---} constrain 
an evening temperature of $1068^{+43}_{-55}$ K and a morning temperature of $889^{+54}_{-65}$ K, implying an evening-to-morning difference of $177^{+65}_{-57}$ K --- a 3$\sigma$ 
difference between the evening and morning limb temperatures consistent with the findings 
with our ATMO best-fit grid model search described above. The rest of the parameters in 
our atmospheric retrievals are all consistent between the morning and the evening limb, 
which suggests that this temperature difference is one of the largest effects defining 
the difference between the morning and evening spectra. In particular, the C/O ratios 
are consistent with each other with the evening limb having 
$\textnormal{C/O} = 0.58^{+0.13}_{-0.16}$ and the morning limb having 
$\textnormal{C/O} = 0.57^{+0.17}_{-0.23}$. Interestingly, while both limbs allow for clouds, 
the cloud-top pressure in the evening limb is much more constrained than the pressure in the 
morning limb --- the cloud-top pressure of the evening cloud being located in our retrievals 
at about $\sim 1-10$ mbar, and clouds in the morning being consistent with a wide range of possibilities. {Notably, the posterior distributions presented in Extended Data Figure \ref{fig:corner_plot} showcase how these cloud properties do not strongly define the morning/evening temperatures, likely stemming from the fact that these two properties are extracted both from the absolute depth difference between the morning and evening spectra \textit{and} the amplitude of the H$_2$O and CO$_2$ features. For example, while the morning cloud-top pressure is consistent with a wide range of values, all those values are allowed within the relatively narrow temperature ranges for the morning/evening temperatures described above. In the case of the evening cloud-top pressure, which is much better defined, some slight correlations are observed with the temperatures, but again within well-constrained temperature values. Finally,} the retrieved metallicity in our fits is consistent with a $\times 10$ solar metallicity %$\times 8.2^{+1.7}_{-1.2}$ solar 
--- again, consistent with our best-fit ATMO models.

\subsection*{General circulation models}
\textbf{Condensate cloud model.} The results from the cloud microphysics model presented here were obtained applying a kinetic cloud formation model (consistent solution of nucleation of different species, growth and evaporation of mixed materials, gravitational settling, mixing, element conservation), coupled to equilibrium gas phase chemistry \cite{Woitke2004quasiclouds,Helling2006dirtyclouds} for 10$\times$ solar element abundances. 1D pressure-temperature profiles extracted from a cloud-free GCM simulation of WASP-39b using ExpeRT/MITgcm \cite{SchneiderEtAl2022ExpeRT} were used as input. The mixing timescale in the model was calculated based on the local vertical velocities from the GCM and scale height (see \cite{HellingEtAl2023ModelGrid}). The mixing timescale was then multiplied by a factor of 100 \cite{parmentier:2013,SamraEtAl2023WASP-96b}.
%following \citet{SamraEtAl2023WASP-96b}.
We note that the GCM simulations are identical to those presented in ref. \cite{CaroneEtAl2023WASP-39b}. The resulting cloud particle number densities, their mixed material compositions and mean particle sizes were used as input to calculate the local cloud opacity with the adapted version of petitRADTRANS \cite{MolliereEtAl2019petitRADTRANS,molliere:2020,alei:2022} using {the Landau, Lifshitz and Looyenga (LLL)} \cite{landau:1960,looyenga:1965} mixing prescription and Mie theory {using the publicly available python code PyMieScatt \cite{sumlin2018retrieving}}. The spectra from {nine} different latitudes {(-86$^\circ$, -68$^\circ$, -45$^\circ$, -23$^\circ$, 0$^\circ$, 23$^\circ$, 45$^\circ$, 68$^\circ$, 86$^\circ$)} were then averaged at the evening and morning terminator, respectively. 

\textbf{Photochemical haze model.} For modeling photochemical hazes, we used the haze model presented in ref. \cite{SteinrueckEtAl2021} in combination with SPARC/MITgcm \cite{ShowmanEtAl2009, KatariaEtAl2013}, which couples wavelength-dependent radiative transfer using the correlated-k method to a dynamical core based on the primitive equations \cite{AdcroftEtAl2004}. All numerical choices in the model are identical to the wavelength-dependent passive model presented in ref. \cite{steinrueck:2023} except that the planet parameters of WASP-39b, a 10$\times$ solar metallicity, and a temperature of the bottom-most layer of 4154~K were assumed. The model treats hazes as passive tracers with constant particle sizes. Hazes are produced at low pressures on the dayside and destroyed at pressures $>$ 0.1~bar. Particle sizes ranging from 1~nm to 1000~nm were considered. The haze production rate then was adjusted to obtain a good match to the total (not terminator-resolved) transit spectrum. Particle sizes of 3~nm and 30~nm resulted in the best match to the transit spectrum. The spectrum shown in Fig. 3 uses a particle size of 30~nm and a haze production rate of $2.5\times 10^{-12}$ kg m$^{-2}$ s$^{-1}$ at the substellar point. The spectra generated with a particle size of 3~nm showed qualitatively similar but smaller differences between morning and evening terminator.

\textbf{Clear-atmosphere equilibrium and transport-induced disequilibrium chemistry models.}
To model equilibrium and disequilibrium chemistry (in particular, transport-induced quenching that occurs when chemical kinetics are coupled to atmospheric transport), WASP-39b was simulated using the Met Office {\sc Unified Model} \cite{Wood2014UM,Mayne2014UM}. This model's dynamical core solves the non-hydrostatic, deep-atmosphere Navier-Stokes equations. Its radiative transfer code \cite{Socrates1996UMRT,Amundsen2014UMRT,Amundsen2016UMRT} solves the two-stream equations using the correlated-k and equivalent extinction methods, including $\rm H_2O$, CO, $\rm CO_2$, $\rm CH_4$, $\rm NH_3$, HCN, Li, Na, K, Rb, Cs and collision-induced absorption due to $\rm H_2–H_2$ and $\rm H_2–He$ as sources of opacity, assuming clear-sky conditions and a 10$\times$ solar metallicity. Two different chemistry schemes were used. One is the chemical equilibrium scheme that computes a local chemical equilibrium using the Gibbs energy minimization, and another is the chemical kinetics scheme which solves the ordinary differential equations describing the evolution of chemical species present in the reduced chemical network of ref. \cite{VenotEtAl2019ReducedScheme} to represent disequilibrium thermochemistry (in the absence of photolysis), as implemented in refs. \cite{DrummondEtAl2020} and \cite{Zamyatina2023Disequilibrium}. To maintain stability, the model employs a vertical sponge, with damping coefficient 0.15, and a diffusion filter in the longitudinal direction, with coefficient $3.83\times10^{-2}$ \cite{Mayne2014UM}.

\textbf{Calculating morning/evening spectra.}
For the photochemical-haze and clear-atmosphere models, we generate transmission spectra of WASP-39b for the morning and evening terminators similarly to the method described in \cite{savel2022no}.   Briefly, we perform absorption-only, ray-striking radiative transfer through the input GCM at two orbital phases. We rotate the GCM by the phase angle at ingress and egress, respectively, in addition to interpolating the GCM onto an equal-altitude grid truncated at approximately 1 bar. For the Met Office {\sc Unified Model}, chemical abundances are taken from the GCM output. For the SPARC/MITgcm photochemical haze model, chemistry is interpolated from FastChem \cite{stock2018fastchem, stock2022fastchem} equilibrium chemistry tables. 

Our opacity sources include $\rm H_2O$ \cite{polyansky2018exomol}, $\rm CH_4$ \cite{yurchenko2014exomol, yurchenko2017hybrid}, $\rm CO$ \cite{li2015rovibrational,somogyi2021calculation}, $\rm CO_2$ \cite{yurchenko2020exomol}, $\rm C_2H_2$ \cite{chubb2020exomol}, and $\rm NH_3$ \cite{al2015marvel,coles2019exomol}. We include the extinction (absorption and scattering out of the beam) from haze particles with Mie theory using \texttt{PyMieScatt} \cite{sumlin2018retrieving}, assuming homogeneous particles with size set by the GCM input and a refractive index of soot \cite{LavvasKoskinen}.

\textbf{A note on 1D vs 3D transit spectra calculations.} We point out that spectra for the condensate cloud model were calculated with a 1D radiative transfer code with the classic mid-transit geometry due to the 1D nature of the microphysics model and the additional complexity of mixed-composition grains. In contrast, spectra for the photochemical haze model and clear-atmosphere equilibrium and disequilibrium chemistry models were calculated with a 3D code that also correctly takes into account the change in geometry during ingress and egress through rotation. We note that in tests comparing 1D and 3D geometries based on the model grid of \cite{RomanEtAl2021}, the amplitude of the terminator differences changed. In some (but not all) cases, this leads to stronger limb asymmetries if the full 3D geometry was considered (Arnold et al., in prep.). This potentially could bring the size of the observed differences into better agreement with observations.

\section*{Extended Data Tables and Figures}

\clearpage

\noindent \textbf{Data Availability}
The raw data from this study is available as part of the Early Release Science Observations (ERS) via the Space Science Telescope Institute's Mikulski Archive for Space Telescopes (\url{https://archive.stsci.edu/}). {All the figures in this manuscript, along with the associated data and code to reproduce them, can be found at \url{https://github.com/nespinoza/wasp39-terminators}. Reduced data along with prior and posterior distributions for our wavelength-dependant \texttt{catwoman} (NE) light curve fits used to obtain the main results of this work can be found at \url{https://stsci.box.com/s/rx7u56zviu3up2p8p34qh3btwop6lgl6}. Reduced data along with prior and posterior distributions for our white-light light curve fit performed for WASP-39~b and described in the Methods section can be found at \url{https://stsci.box.com/s/wet5xmacrk26ughr8y2j8wpyjdsumco1}. Both datasets contain human-readable outputs, and are packaged to be explored using the \texttt{juliet} software library, which is publicly available at \url{https://github.com/nespinoza/juliet}.}
\\

\noindent \textbf{Code Availability}\\
Light curves were fitted using \texttt{juliet} 
(\url{https://github.com/nespinoza/juliet}), \texttt{batman} (\url{https://github.com/lkreidberg/batman}), \texttt{catwoman} (\url{https://github.com/KathrynJones1/catwoman}) and \texttt{Tiberius} (\url{https://github.com/JamesKirk11/Tiberius}), all of which are publicly available. \\

\vskip 1 cm
\backmatter

\noindent \textbf{Acknowledgements}
\noindent This work is based on observations made with the NASA/ESA/CSA James Webb Space Telescope. The data were obtained from the Mikulski Archive for Space Telescopes at the Space Telescope Science Institute, which is operated by the Association of Universities for Research in Astronomy, Inc., under NASA contract NAS 5-03127 for JWST. These observations are associated with program \#1366. MS acknowledges
support from the 51 Pegasi b Fellowship funded by the Heising-Simons
Foundation.\\

% Add funding acknoledgments.

\noindent \textbf{Author Contributions}\\
NE led the main analyses and the writing of the paper.
NE performed atmospheric retrievals, which were discussed with RM and LW.
MS led the compilation of the GCM forward modelling effort, and led the writing of the GCM section of the paper. JK, MM and ZR performed lightcurve fits and provided data analysis expertise and feedback to the project as a whole.
MZ, DC and NM performed clear-atmosphere equilibrium and transport-induced disequilibrium chemistry simulations. LC, DL, DS and SK performed the cloud forward model simulations. AS post-processed the GCMs. MLP, EK, ER, MR, AS, MM, JK, MZ, DC, LC, JB, LD, SMT, EP, LM, BVR, AC, NA, KM, ZR, NC and VP provided comments to the manuscript. EM \& AC led the compilation and analysis of the detector-level data, as well as expertise 
on the data reduction and analysis. RB performed the analysis to obtain the stellar density for WASP-39~b. All co-authors read and agreed with the conclusions of the manuscript.\\ 
% Need to add: instrument/dataset expertise (e.g., Erin May, Aaryn Carter), forward models, GCMs,

\noindent \textbf{Competing interests} The authors declare no competing interests.\\

\noindent\textbf{Additional information}\newline
\textbf{Correspondence and requests for materials} should be addressed to \href{Nestor Espinoza}{mailto:nespinoza@stsci.edu}.\newline
\textbf{Reprints and permissions information} is available at \url{www.nature.com/reprints}.

%\bibliography{sn-bibliography}% common bib file
%% if required, the content of .bbl file can be included here once bbl is generated
%%\input sn-article.bbl

\end{document}